\pdfoutput=1 % to force arxiv to use pdflatex
\documentclass[aps,prl,twocolumn,groupedaddress,amsmath]{revtex4-1}

\usepackage{color}
\usepackage{graphicx}
\usepackage{amssymb}
\usepackage{bm}
\usepackage{amsmath,amsfonts,latexsym}

\usepackage{braket} % For bra-ket notation

\usepackage{comment}

\usepackage{bm} % allows bold greek letters

\begin{document}

\title{Noise correlations in the expansion of an interacting 1D Bose gas from a regular array}

\author{Austen Lamacraft} 
\affiliation{Department of Physics, University of Virginia,
Charlottesville, Virginia 22904-4714 USA}
\date{\today}
\email{austen@virginia.edu}
 
\date{\today}

\begin{abstract}

We consider the one dimensional expansion of a system of interacting bosons, starting from a regular array. Without interactions the familiar Hanbury Brown and Twiss effect for bosons gives rise to a series of peaks in the density-density correlations of the expanded system. Infinitely repulsive particles likewise give a series of dips, a signature of the underlying description in terms of free fermions. In the intermediate case of finite interaction the noise correlations consist of a set of Fano resonance lineshapes, with an asymmetry parameter determined by the scattering phase shift of a pair of particles, and a width depending on the initial momentum spread of the particles.

\end{abstract}

\maketitle

The Hanbury Brown and Twiss (HBT) effect \cite{brown:1956} is a fundamental signature of quantum statistics appearing in quantum optics, atomic and mesoscopic physics, and nuclear collisions \cite{kleppner2008,schellekens2005,henny1999,oliver1999,baym1998}. It is most dramatically manifested as an interference effect in the intensity correlations due to two or more incoherent sources, with a sign depending on the statistics of particles: positive correlations for bosons; negative for fermions. 

%In Ref.~\cite{altman2004} it was pointed out that intensity correlations could be extracted from the noise present in the time-of-flight images ubiquitous in ultracold atomic physics. 

In most known instances of the HBT effect interactions between particles do not play a significant role, either because these effects are weak or due to the spatial separation of the sources. In this Letter we consider the one-dimensional expansion of a system of particles, where strong interaction effects are unavoidable. Indeed, in 1D the trajectories giving rise to the HBT effect must cross. 

\begin{figure}
	\centering
		\def\svgwidth{1.1\columnwidth}
		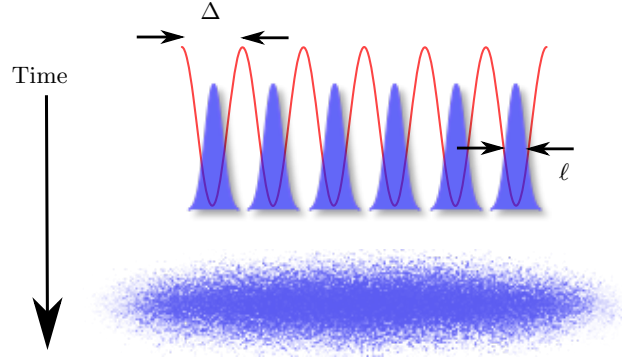
	\caption{1D expansion of atoms from an optical lattice. Noise correlations will be present in an absorption image of the expanded cloud.}
	\label{fig:array}
\end{figure}

The situation that we will consider is illustrated in Fig.~\ref{fig:array}. Particles are initially confined to a regular 1D lattice of spacing $\Delta$, with one particle per site. At time $t=0$ the lattice potential is removed, though the potential restricting the particles' motion to one dimension remains. We are concerned with the density correlations present after some time $t$, when the system has expanded to many times its original size (analogous to the `far field' limit in optics). Thus we have in mind a 1D version of the experiment of Ref.~\cite{folling:2005}, in which noise correlations were measured in the expansion of a 3D atomic Mott insulating state from an optical lattice. A recent experiment demonstrated the preparation of such a 1D state in a slightly different context \cite{trotzky:2011}.

\begin{figure}
	\centering
		\def\svgwidth{1\columnwidth}
		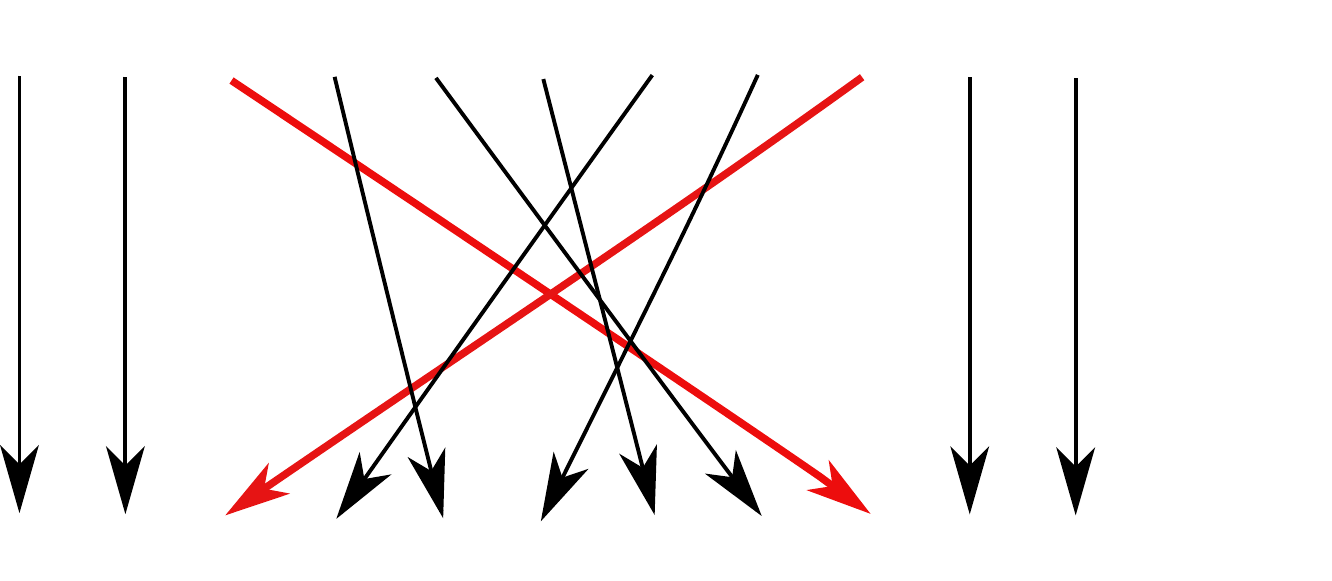
		\def\svgwidth{1\columnwidth}
		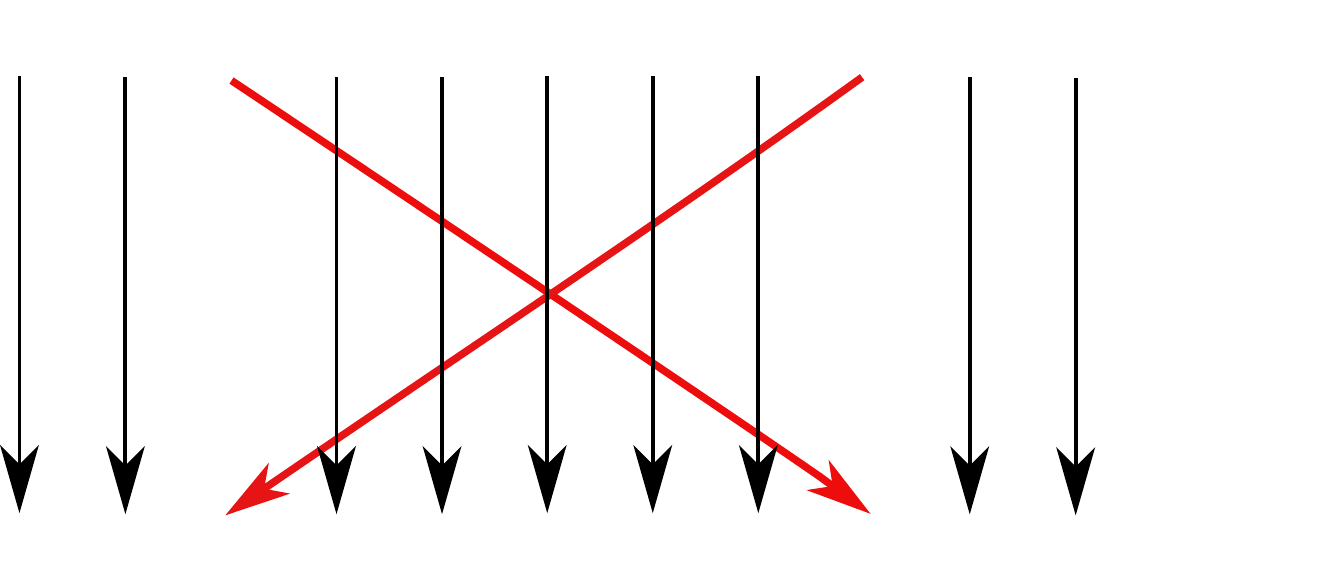
	\caption{(Top) Diagram representing a term that survives integration over $x_{j}$, $j\neq 1,2$. At the top are the positions $\left\{y_{\sigma_{1}(i)}\right\}$, and at the bottom are $\left\{\tilde y_{\sigma_{2}(i)}\right\}$. The thick red lines correspond to the two coordinates $x_{1}$ and $x_{2}$ that are not integrated over in the two-body density matrix Eq.~\eqref{HBTLL_FullArray}. (Bottom) Usual contribution to the noninteracting HBT effect.}
	\label{fig:permutation}
\end{figure}

To introduce some ideas and notation we briefly describe the familiar HBT effect in this setting. We assume Gaussian initial wavefunctions corresponding to harmonic oscillator length $\ell=\sqrt{\hbar/m\omega}$, $\varphi_{\alpha}(y)= \frac{1}{(\pi \ell^{2})^{1/4}} \exp\left[-\frac{(y-\alpha\Delta )^{2}}{2\ell^{2}}\right]$.
The overlap $e^{-\Delta^{2}/4\ell^{2}}$ between neighboring sites is assumed to be negligible. After a period $t$ of free evolution these wavefunctions have the form
\begin{equation}
	\label{HBTLL_freeGauss}
	\varphi_{\alpha}(x;t\gg \ell^{2})\to\sqrt{\frac{\ell}{i\sqrt{\pi} t}}\exp\left[\left(\frac{i}{2t}-\frac{\ell^{2}}{2t^{2}}\right)\left(x-\alpha\Delta\right)^{2}\right].
\end{equation}
(Where we have set $\hbar=m=1$)
If we consider a pair of identical particles on sites $\alpha$ and $\alpha+1$, the two-particle wavefunction is $\Psi_{2}(x_{1},x_{2};t)=\frac{1}{\sqrt{2}}\left[\varphi_{\alpha}(x_{1};t)\varphi_{\alpha+1}(x_{2};t)\pm\varphi_{\alpha}(x_{2};t)\varphi_{\alpha+1}(x_{1};t)\right]$, with $\pm$ for bosons and fermions respectively. The corresponding probability density is then
\begin{multline}
	\label{HBTLL_2part}
	|\Psi_{2}(x_{1},x_{2};t)|^{2}\to\frac{\ell^{2}}{\pi t^{2}}e^{-\ell^{2}\left(\xi_{1}^{2}+\xi_{2}^{2}\right)}\left[1\pm\cos\left(\left[\xi_{1}-\xi_{2}\right]\Delta\right)\right],
\end{multline}
where the variables $\xi_{1,2}=x_{1,2}/t$ correspond to the velocities of the two particles. The oscillatory second term describes the HBT effect, with a sign dependent on the statistics of the particles. For an array of $N$ particles the density-density correlation function develops peaks due to the contributions of higher harmonics arising from pairs of particles separated by multiples of $\Delta$
%
%\xrightarrow[N\to\infty]{}
\begin{multline}
	\label{HBTLL_FullArray}
	\mathcal{C}(x_{1},x_{2};t)\equiv\int dx_{3}\cdots dx_{N} |\Psi_{N}(x_{1},x_{2},\ldots,x_{N};t)|^{2}.\\
	\to\frac{\ell^{2}}{\pi t^{2}}e^{-\ell^{2}\left(\xi_{1}^{2}+\xi_{2}^{2}\right)}
	\left[1\pm\frac{2\pi}{N}\sum_{n=-\infty}^{\infty}\delta(\Delta\left[\xi_{1}-\xi_{2}\right]-2\pi n)\right]
\end{multline}
In a trajectory picture the HBT effect arises as a cross-term between trajectories that do and do not exchange pairs of particles (see Fig.~\ref{fig:permutation}, bottom)

We turn now to the central subject of this paper: the HBT effect in the presence of interactions between the particles. We assume that the evolution of the system for $t>0$ is governed by the $N$-particle Hamiltonian
\begin{equation}
	\label{HBTLL_Ham}
	H=-\frac{1}{2}\sum_{i=1}^{N}\frac{\partial^{2}}{\partial x_{i}^{2}}+c\sum_{i<j}\delta(x_{i}-x_{j}).
\end{equation}
\begin{figure}
	\centering
		\def\svgwidth{1.1\columnwidth}
		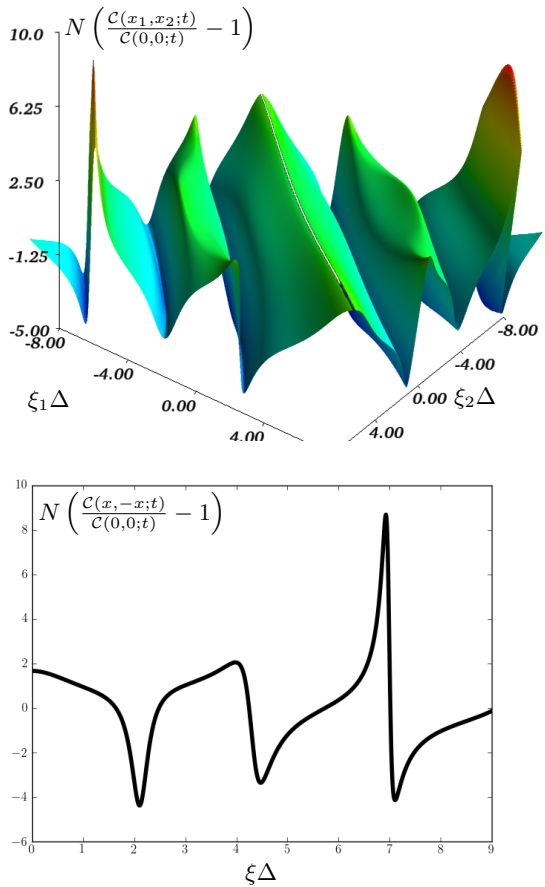
	\caption{(Top) Normalized correlation function $N\left(\frac{\mathcal{C}(x_{1},x_{2};t)}{\mathcal{C}(0,0;t)}-1\right)$ for $c\Delta=2$, $\ell/\Delta=0.2$. (Bottom) A slice with $x_{1}=-x_{2}$ for the same parameters, showing the evolution of the Fano asymmetry between successive peaks.}
	\label{fig:fano}
\end{figure}
The $c\to 0$ and  $c\to\infty$ limits can be described in terms of free bosons and free fermions, respectively. The density-density correlations reflect this, corresponding to the plus sign in Eq.~\eqref{HBTLL_FullArray} in the former case and the minus sign in the latter. %Is there a simple picture of the crossover between these two limits? 
Our main result, valid when $e^{-2c\Delta}\ll 1$, is that in the crossover regime the density-density correlations consist not of a series of symmetric peaks or dips but rather of \emph{Fano lineshapes} (see Fig.~\ref{fig:fano})
\begin{equation}
	\label{expansion_fano} 
	\frac{\left[q_{n}\Gamma_{n}/2+\left(\varepsilon-\eta_{n}\right)\right]^{2}}{\Gamma_{n}^{2}/4+\left(\varepsilon-\eta_{n}\right)^{2}}
\end{equation}
where $\varepsilon=\Delta\left(\xi_{1}-\xi_{2}\right)-2\pi n$ represents the deviation from the $n^{\text{th}}$ peak. The asymmetry parameter $q_{n}$ is expressed in terms of the two particle scattering matrix 
\begin{equation}
	\label{expansion_Smatrix}
	S(k)=-\frac{c-ik}{c+ik},
\end{equation}
by the relation 
\begin{equation}
	\label{HBTLL_asym}
	\arg S(2\pi n/\Delta)=2q_{n}/(q_{n}^{2}-1).
\end{equation}
This illustrates the evolution from $q_{n}=\infty$ for free bosons (resonance lineshape) to $q_{n}\to 0$ as $c\to\infty$ (antiresonance). The asymmetry of the lineshape is the first qualitative feature of the crossover regime. The second is the finite width $\Gamma_{n}$, for which we give the explicit form below, and which vanishes in the two limits.

The surprising simplicity of our result is a consequence of the integrability of the Hamiltonian Eq.~\eqref{HBTLL_Ham} \cite{lieb:1963}. The $N$-particle scattering it describes is \emph{nondiffractive}, consisting of pairwise scattering that either preserves or exchanges the momenta of the scattering particles. A remarkable consequence is that the time dependence of the $N$-particle propagator describing the amplitude for particles at $y_{1},\ldots,y_{N}$ to arrive at $x_{1},\ldots x_{N}$ after time $t$ can be written explicitly for $c>0$ as \cite{tracy:2008}

\begin{widetext}
	\begin{equation}
		\label{expansion_TWform}
		\mathcal{G}_{N}(x_{1},x_{2},\ldots,x_{N}|y_{1},y_{2},\ldots,y_{N};t)=\sum_{\sigma\in \mathcal{S}_{N}}\int\cdots\int A_{\sigma}\prod_{j=1}^{N}e^{ik_{\sigma(j)}(x_{j}-y_{\sigma(j)})}e^{-\frac{it}{2}\sum_{j}k_{j}^{2}}\frac{dk_{1}}{2\pi}\cdots \frac{dk_{N}}{2\pi}
	\end{equation}	
\end{widetext}
where $\mathcal{S}_{N}$ denotes the symmetric group of degree $N$, and
\begin{equation}
	\label{expansion_Adef}
	A_{\sigma}=\prod\left\{S(k_{\sigma(\alpha)}-k_{\sigma(\beta)}): x_{\alpha}<x_{\beta} \text{ but } y_{\sigma(\alpha)}>y_{\sigma(\beta)} \right\}.
\end{equation}
%
%\texttt{Not just a function of $\sigma$ of course!}
To verify Eq.~\eqref{expansion_TWform} one should first observe that it satisfies the boundary condition $\left(\frac{\partial}{\partial x_{i}}-\frac{\partial}{\partial x_{j}} \right)\mathcal{G}_{N}|_{x_{i}=x_{j}}=c\, \mathcal{G}_{N}|_{x_{i}=x_{j}}$ imposed by the interaction. Next we must check that the initial condition $\mathcal{G}_{N}(\mathbf{x}|\mathbf{y};0)=\sum_{\sigma}\prod_{i}\delta(x_{i}-y_{\sigma(j)})$ is obeyed. This follows from the fact that the integral 
\begin{equation}
	\label{HBTLL_t0}
	\int\cdots\int A_{\sigma}\prod_{j=1}^{N}e^{ik_{\sigma(j)}(x_{j}-y_{\sigma(j)})}\frac{dk_{1}}{2\pi}\cdots \frac{dk_{N}}{2\pi}
\end{equation}
is nonzero only for $A_{\sigma}=1$ i.e. when the $\left\{x_{i}\right\}$ are in the same order as the $\left\{y_{i}\right\}$. This in turn is a consequence of the following Golden Rule that we will use repeatedly for integrals of this type \footnote{`Never impose on others what you would not choose for yourself.' -- Confucius} : \emph{a particle moving to the left (right) must be overtaken by another particle moving to the left (right)}. In the present case the Golden Rule restricts us to $A_{\sigma}=1$, from which the product of $\delta$-functions follows. 

To understand the origin of the Golden Rule, consider the integral over $k_{\sigma(j)}$ in Eq.~\eqref{HBTLL_t0}. The result can be viewed as the Fourier transform of the product of factors in $A_{\sigma}$ involving $k_{\sigma(j)}$, evaluated at $x_{j}-y_{\sigma(j)}$, which is thus the convolution of the Fourier transforms of these factors. Because $S(k)$ is holomorphic in the lower half plane for $c>0$, its Fourier transform is supported in $[0,\infty)$. Thus for $x_{j}-y_{\sigma(j)}>0$ we must have at least one factor $S(k_{\sigma(j)}-k_{\sigma(i)})$ where $k_{\sigma(j)}$ appears first (particle is overtaken moving to the right). Likewise for $x_{j}-y_{\sigma(j)}<0$ we must have at least one factor $S(k_{\sigma(k)}-k_{\sigma(j)})$ where $k_{\sigma(j)}$ appears second (particle is overtaken moving to the left).

The time evolution of our array can be found by convolving the propagator with the Gaussian initial wavepackets 
\begin{equation}
	\label{HBTLL_evol}
	\Psi_{N}(\mathbf{x};t)=\frac{1}{\sqrt{N!}}\int \mathcal{G}_{N}(\mathbf{x},\mathbf{y};t)\prod_{j}\varphi_{j}(y_{i}) d \mathbf{y}
\end{equation}
The utility of this expression would seem to be hampered by the momentum integrals and the sum over permutations in Eq.~\eqref{expansion_TWform}. However, the former may be evaluated in the stationary phase approximation at long times
\begin{equation}
	\label{expansion_FarField}
	\mathcal{G}_{N}(\mathbf{x}|\mathbf{y};t)\to\left(\frac{1}{2\pi i t}\right)^{N/2}\sum_{\sigma\in \mathcal{S}_{N}} A'_{\sigma}\prod_{j=1}^{N}e^{i\left(\frac{t}{2}\xi_{j}^{2}-\xi_{j}y_{\sigma(j)}\right)},
\end{equation}
where again we have used the variables $\xi_{j}=x_{j}/t$, and the stationary phase integral assumes that these are order one in the long time limit. In the above $A'_{\sigma}$ denotes 
\begin{equation}
	\label{expansion_Adef2}
	A'_{\sigma}=\prod\left\{S(\xi_{\alpha}-\xi_{\beta}): x_{\alpha}<x_{\beta} \text{ but } y_{\sigma(\alpha)}>y_{\sigma(\beta)} \right\}.
\end{equation}
To evaluate the probability distribution we require the `forward and back' propagator
\begin{widetext}
	\begin{equation}
		\label{expansion_fwdbk}
		\mathcal{G}_{N}(\mathbf{x}|\mathbf{y};t)\mathcal{G}^{*}_{N}(\mathbf{x}|\mathbf{\tilde y};t)\to
		\left(\frac{1}{2\pi t}\right)^{N}\sum_{\sigma_{1},\sigma_{2}\in \mathcal{S}_{N}}A'_{\sigma_{1}}A'^{*}_{\sigma_{2}}\prod_{j}e^{-i\xi_{j}\left(y_{\sigma_{1}(j)}-\tilde y_{\sigma_{2}(j)}\right)}.
	\end{equation}
In this expression the scattering phases have the explicit form
	\begin{equation}
		\label{expansion_SqScat}
		A'_{\sigma_{1}}A'^{*}_{\sigma_{2}}=\prod \left\{S(\xi_{\alpha}-\xi_{\beta}):\sigma_{1}(\alpha)>\sigma_{1}(\beta) \text{ but } \sigma_{2}(\alpha)<\sigma_{2}(\beta) \right\},
	\end{equation}
\end{widetext}
Unlike the individual $A'_{\sigma}$, we see that the form of the product does not depend upon the ordering of the $\left\{x_{j}\right\}$. Since we need to integrate over all but two of the $\left\{x_{j}\right\}$ to find the density correlation function (see Eq.~\eqref{HBTLL_FullArray}), this fact is tremendously useful, as it tells us that the integrals have the same form as Eq.~\eqref{HBTLL_t0}, and allows us to apply the Golden Rule. The only non-trivial terms (i.e. without $\sigma_{1}(\alpha)=\sigma_{2}(\alpha)$ for all $\alpha$) are of the form illustrated in Fig.~\ref{fig:permutation} (top). $x_{1}$ and $x_{2}$ are exempted from the Golden Rule and correspond to the only particles not overtaken. Thus we must have $\sigma_{1}(1)=\sigma_{2}(2)$ and $\sigma_{2}(1)=\sigma_{1}(2)$

Despite this simplification there would still seem to be a great many terms to sum in Eq.~\eqref{expansion_fwdbk}. We will now show that the remaining terms can be grouped according to the order of their contribution in the parameter $e^{-2c\Delta}$, with the lower powers amenable to explicit evaluation. Since the parameter $c\Delta$ is the same as the usual Lieb--Liniger parameter $\gamma\equiv c/n$, 	with the density $n=\Delta^{-1}$, the use of $e^{-2c\Delta}$ as a small parameter is not too restrictive.

Let us first consider the terms that give rise to the usual HBT effect in the case of noninteracting particles (Fig.~\ref{fig:permutation}, bottom). Each of the $x_{\alpha}$ with $\sigma_{1}(\alpha)=\sigma_{2}(\alpha)$ lying between $\sigma_{1}(1)=\sigma_{2}(2)$ and $\sigma_{2}(1)=\sigma_{1}(2)$ brings a factor $S(\xi_{2}-\xi_{\alpha})S(\xi_{\alpha}-\xi_{1})$ if $\sigma_{1}(1)<\sigma_{1}(2)$ and $S(\xi_{1}-\xi_{\alpha})S(\xi_{\alpha}-\xi_{2})$ if $\sigma_{1}(2)<\sigma_{1}(1)$. After integrating over the $\left\{x_{\alpha}:\alpha\neq 1,2\right\}$ and convolving with the Gaussian wavepackets Eq.~\eqref{HBTLL_evol}, we can sum all such contributions in a geometric series to give
\begin{multline}
	\label{expansion_final}
	\mathcal{C}(x_{1},x_{2}:t)\to
	\frac{\ell^{2}}{\pi t^{2}}e^{-\ell^{2}\left(\xi_{1}^{2}+\xi_{2}^{2}\right)}\\
	\times\left[1+\frac{2}{N}\text{Re}\left(\frac{S(\xi_{2}-\xi_{1}) e^{i\Delta(\xi_{1}-\xi_{2})}}{1-e^{i\Delta(\xi_{1}-\xi_{2})}\zeta(\xi_{1},\xi_{2})}\right)\right],
\end{multline}
which generalizes Eq.~\eqref{HBTLL_FullArray} to the interacting case. In Eq.~\eqref{expansion_final} we have defined the function
\begin{multline*}
	%\label{expansion_dampfactor}
	\zeta(\xi_{1},\xi_{2})=1-2\sqrt{\pi}c\ell S(\xi_{2}-\xi_{1}-ic)\\
\times\left[e^{\ell^{2}(c-i\xi_{1})^{2}}\text{erfc}(\ell\left[c-i\xi_{1}\right])+e^{\ell^{2}(c+i\xi_{2})^{2}}\text{erfc}(\ell\left[c+i\xi_{2}\right]) \right],
\end{multline*}
where $\text{erfc}(x)$ is the complementary error function $\text{erfc}(x)=\frac{2}{\sqrt{\pi}}\int_{x}^{\infty}e^{-t^{2}}dt$. Eq.~\eqref{expansion_final} is shown in Fig.~\ref{fig:fano}

In the limit that $\zeta(\xi_{1},\xi_{2})$ is close to unity, Eq.~\eqref{expansion_final} can be interpreted as a series of Fano lineshapes Eq.~\eqref{expansion_fano} with $\eta_{n}=-\text{Im}\,\zeta$, $\Gamma_{n}=2(1-\text{Re}\,\zeta)>0$, and $q_{n}$ as given in Eq.~\eqref{HBTLL_asym}. Fig.~\ref{fig:fano} (bottom) illustrates the evolution of $q_{n}$ between successive peaks from smaller (close to antiresonance) to larger values.

%\texttt{At what values??}

The physical origin of the asymmetry $q_{n}$ lies in the scattering phase of particles 1 and 2 with each other, while the width $\Gamma_{n}$ arises from the collisions of these particles with those that they pass, whose momentum has a Gaussian distribution and gives rise to a distribution of scattering phases. $\Gamma_{n}$ vanishes in the limits $c\to 0$ and $c\to \infty$, but also when $\ell\to 0$. In the last case this is a consequence of the typical momenta of the particles becoming large (except for particles 1 and 2 whose momenta are fixed by $x_{1}$ and $x_{2}$) and the scattering phase for their collisions approaching zero (see Eq.~\eqref{expansion_Smatrix}).

\begin{figure}
	\centering
		\includegraphics[width=0.5\columnwidth]{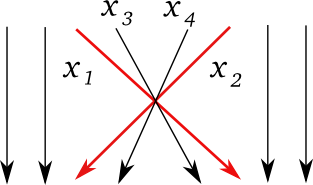}
	\caption{Simplest `new' contribution, smaller by $e^{-2c\Delta}$.}
	\label{fig:permutation_new}
\end{figure}
Let us now show that the remaining contributions are small in the parameter $e^{-2c\Delta}$. Consider the first `non HBT' diagram shown in Fig.~\ref{fig:permutation_new}. Evaluating this diagram gives the contribution
\begin{multline}
	\label{expansion_newave}
		\frac{16 c^{2}\ell^{4}}{t^{2}}e^{-2c\Delta}e^{4i\Delta(\xi_{1}-\xi_{2})}e^{-\ell^{2}(\xi_{1}^{2}+\xi_{2}^{2})}\\
		\times\exp\left[\ell^{2}((c-i\xi_{1})^{2}+(c+i\xi_{2})^{2})\right]\\
		\times S(\xi_{2}-\xi_{1})S(\xi_{2}-\xi_{1}-ic)^{2}S(\xi_{2}-\xi_{1}-2ic)		
\end{multline}
where, as always, we ignore the overlap $e^{-\Delta^{2}/4\ell^{2}}$ between neighboring sites. The two exponential factors $e^{-c\Delta}$ arise from the pole in the upper half plane of $x_{3}$ coming from the $S(\xi_{3}-\xi_{1})$ factor, and from the pole in the lower half plane of $x_{4}$ coming from the $S(\xi_{2}-\xi_{4})$ factor. In the same way, one can show that the power of $e^{-c\Delta}$ appearing in a contribution is at least twice the total number of moves to the right (or to the left).

%\texttt{What about attractive case: holomorphic in UHP}

In conclusion, we have shown that the HBT effect of interacting particles in one dimension has a number of interesting features that distinguish it from the noninteracting problem, most notably an asymmetry and finite width in the peaks of the density-density correlation function of the expanded system. The calculation hinges upon the integrability of the 1D Bose gas, and indeed appears to depend essentially upon $c>0$ for the form of the propagator Eq.~\eqref{expansion_TWform} to be valid.

The author would like to acknowledge the support of the NSF through grant DMR-0846788 and Research Corporation through a Cottrell Scholar award, and thanks Tom Jackson for a useful conversation.

%\bibliography{../Literature/exp}  

\begin{thebibliography}{11}%
\makeatletter
\providecommand \@ifxundefined [1]{%
 \@ifx{#1\undefined}
}%
\providecommand \@ifnum [1]{%
 \ifnum #1\expandafter \@firstoftwo
 \else \expandafter \@secondoftwo
 \fi
}%
\providecommand \@ifx [1]{%
 \ifx #1\expandafter \@firstoftwo
 \else \expandafter \@secondoftwo
 \fi
}%
\providecommand \natexlab [1]{#1}%
\providecommand \enquote  [1]{``#1''}%
\providecommand \bibnamefont  [1]{#1}%
\providecommand \bibfnamefont [1]{#1}%
\providecommand \citenamefont [1]{#1}%
\providecommand \href@noop [0]{\@secondoftwo}%
\providecommand \href [0]{\begingroup \@sanitize@url \@href}%
\providecommand \@href[1]{\@@startlink{#1}\@@href}%
\providecommand \@@href[1]{\endgroup#1\@@endlink}%
\providecommand \@sanitize@url [0]{\catcode `\\12\catcode `\$12\catcode
  `\&12\catcode `\#12\catcode `\^12\catcode `\_12\catcode `\%12\relax}%
\providecommand \@@startlink[1]{}%
\providecommand \@@endlink[0]{}%
\providecommand \url  [0]{\begingroup\@sanitize@url \@url }%
\providecommand \@url [1]{\endgroup\@href {#1}{\urlprefix }}%
\providecommand \urlprefix  [0]{URL }%
\providecommand \Eprint [0]{\href }%
\@ifxundefined \urlstyle {%
  \providecommand \doi  [0]{\begingroup \@sanitize@url \@doi}%
  \providecommand \@doi [1]{\endgroup \@@startlink {\doibase
  #1}doi:\discretionary {}{}{}#1\@@endlink }%
}{%
  \providecommand \doi  [0]{doi:\discretionary{}{}{}\begingroup
  \urlstyle{rm}\Url }%
}%
\providecommand \doibase [0]{http://dx.doi.org/}%
\providecommand \Doi [0]{\begingroup \@sanitize@url \@Doi }%
\providecommand \@Doi  [1]{\endgroup\@@startlink{\doibase#1}\@@Doi}%
\providecommand \@@Doi [1]{#1\@@endlink}%
\providecommand \selectlanguage [0]{\@gobble}%
\providecommand \bibinfo  [0]{\@secondoftwo}%
\providecommand \bibfield  [0]{\@secondoftwo}%
\providecommand \translation [1]{[#1]}%
\providecommand \BibitemOpen [0]{}%
\providecommand \bibitemStop [0]{}%
\providecommand \bibitemNoStop [0]{.\EOS\space}%
\providecommand \EOS [0]{\spacefactor3000\relax}%
\providecommand \BibitemShut  [1]{\csname bibitem#1\endcsname}%
%</preamble>
\bibitem [{\citenamefont {Brown}\ and\ \citenamefont
  {Twiss}(1956)}]{brown:1956}%
  \BibitemOpen
  \bibfield  {author} {\bibinfo {author} {\bibfnamefont {R.}~\bibnamefont
  {Brown}}\ and\ \bibinfo {author} {\bibfnamefont {R.}~\bibnamefont {Twiss}},\
  }\href@noop {} {\bibfield  {journal} {\bibinfo  {journal} {Nature},\ }\textbf
  {\bibinfo {volume} {177}},\ \bibinfo {pages} {27} (\bibinfo {year}
  {1956})}\BibitemShut {NoStop}%
\bibitem [{\citenamefont {Kleppner}(2008)}]{kleppner2008}%
  \BibitemOpen
  \bibfield  {author} {\bibinfo {author} {\bibfnamefont {D.}~\bibnamefont
  {Kleppner}},\ }\href@noop {} {\bibfield  {journal} {\bibinfo  {journal}
  {Physics Today},\ }\textbf {\bibinfo {volume} {61}},\ \bibinfo {pages} {8}
  (\bibinfo {year} {2008})}\BibitemShut {NoStop}%
\bibitem [{\citenamefont {Schellekens}\ \emph {et~al.}(2005)\citenamefont
  {Schellekens}, \citenamefont {Hoppeler}, \citenamefont {Perrin},
  \citenamefont {Gomes}, \citenamefont {Boiron}, \citenamefont {Aspect},\ and\
  \citenamefont {Westbrook}}]{schellekens2005}%
  \BibitemOpen
  \bibfield  {author} {\bibinfo {author} {\bibfnamefont {M.}~\bibnamefont
  {Schellekens}}, \bibinfo {author} {\bibfnamefont {R.}~\bibnamefont
  {Hoppeler}}, \bibinfo {author} {\bibfnamefont {A.}~\bibnamefont {Perrin}},
  \bibinfo {author} {\bibfnamefont {J.}~\bibnamefont {Gomes}}, \bibinfo
  {author} {\bibfnamefont {D.}~\bibnamefont {Boiron}}, \bibinfo {author}
  {\bibfnamefont {A.}~\bibnamefont {Aspect}}, \ and\ \bibinfo {author}
  {\bibfnamefont {C.}~\bibnamefont {Westbrook}},\ }\href@noop {} {\bibfield
  {journal} {\bibinfo  {journal} {Science},\ }\textbf {\bibinfo {volume}
  {310}},\ \bibinfo {pages} {648} (\bibinfo {year} {2005})}\BibitemShut
  {NoStop}%
\bibitem [{\citenamefont {Henny}\ \emph {et~al.}(1999)\citenamefont {Henny},
  \citenamefont {Oberholzer}, \citenamefont {Strunk}, \citenamefont {Heinzel},
  \citenamefont {Ensslin}, \citenamefont {Holland},\ and\ \citenamefont
  {Sch\"onenberger}}]{henny1999}%
  \BibitemOpen
  \bibfield  {author} {\bibinfo {author} {\bibfnamefont {M.}~\bibnamefont
  {Henny}}, \bibinfo {author} {\bibfnamefont {S.}~\bibnamefont {Oberholzer}},
  \bibinfo {author} {\bibfnamefont {C.}~\bibnamefont {Strunk}}, \bibinfo
  {author} {\bibfnamefont {T.}~\bibnamefont {Heinzel}}, \bibinfo {author}
  {\bibfnamefont {K.}~\bibnamefont {Ensslin}}, \bibinfo {author} {\bibfnamefont
  {M.}~\bibnamefont {Holland}}, \ and\ \bibinfo {author} {\bibfnamefont
  {C.}~\bibnamefont {Sch\"onenberger}},\ }\href@noop {} {\bibfield  {journal}
  {\bibinfo  {journal} {Science},\ }\textbf {\bibinfo {volume} {284}},\
  \bibinfo {pages} {296} (\bibinfo {year} {1999})}\BibitemShut {NoStop}%
\bibitem [{\citenamefont {Oliver}\ \emph {et~al.}(1999)\citenamefont {Oliver},
  \citenamefont {Kim}, \citenamefont {Liu},\ and\ \citenamefont
  {Yamamoto}}]{oliver1999}%
  \BibitemOpen
  \bibfield  {author} {\bibinfo {author} {\bibfnamefont {W.}~\bibnamefont
  {Oliver}}, \bibinfo {author} {\bibfnamefont {J.}~\bibnamefont {Kim}},
  \bibinfo {author} {\bibfnamefont {R.}~\bibnamefont {Liu}}, \ and\ \bibinfo
  {author} {\bibfnamefont {Y.}~\bibnamefont {Yamamoto}},\ }\href@noop {}
  {\bibfield  {journal} {\bibinfo  {journal} {Science},\ }\textbf {\bibinfo
  {volume} {284}},\ \bibinfo {pages} {299} (\bibinfo {year}
  {1999})}\BibitemShut {NoStop}%
\bibitem [{\citenamefont {Baym}(1998)}]{baym1998}%
  \BibitemOpen
  \bibfield  {author} {\bibinfo {author} {\bibfnamefont {G.}~\bibnamefont
  {Baym}},\ }\href@noop {} {\bibfield  {journal} {\bibinfo  {journal} {Acta
  Physica Polonica B},\ }\textbf {\bibinfo {volume} {29}},\ \bibinfo {pages}
  {1839} (\bibinfo {year} {1998})}\BibitemShut {NoStop}%
\bibitem [{\citenamefont {F\"olling}\ \emph {et~al.}(2005)\citenamefont
  {F\"olling}, \citenamefont {Gerbier}, \citenamefont {Widera}, \citenamefont
  {Mandel}, \citenamefont {Gericke},\ and\ \citenamefont
  {Bloch}}]{folling:2005}%
  \BibitemOpen
  \bibfield  {author} {\bibinfo {author} {\bibfnamefont {S.}~\bibnamefont
  {F\"olling}}, \bibinfo {author} {\bibfnamefont {F.}~\bibnamefont {Gerbier}},
  \bibinfo {author} {\bibfnamefont {A.}~\bibnamefont {Widera}}, \bibinfo
  {author} {\bibfnamefont {O.}~\bibnamefont {Mandel}}, \bibinfo {author}
  {\bibfnamefont {T.}~\bibnamefont {Gericke}}, \ and\ \bibinfo {author}
  {\bibfnamefont {I.}~\bibnamefont {Bloch}},\ }\href@noop {} {\bibfield
  {journal} {\bibinfo  {journal} {Nature},\ }\textbf {\bibinfo {volume}
  {434}},\ \bibinfo {pages} {481} (\bibinfo {year} {2005})}\BibitemShut
  {NoStop}%
\bibitem [{\citenamefont {Trotzky}\ \emph {et~al.}(2011)\citenamefont
  {Trotzky}, \citenamefont {Chen}, \citenamefont {Flesch}, \citenamefont
  {McCulloch}, \citenamefont {Schollw\"ock}, \citenamefont {Eisert},\ and\
  \citenamefont {Bloch}}]{trotzky:2011}%
  \BibitemOpen
  \bibfield  {author} {\bibinfo {author} {\bibfnamefont {S.}~\bibnamefont
  {Trotzky}}, \bibinfo {author} {\bibfnamefont {Y.}~\bibnamefont {Chen}},
  \bibinfo {author} {\bibfnamefont {A.}~\bibnamefont {Flesch}}, \bibinfo
  {author} {\bibfnamefont {I.}~\bibnamefont {McCulloch}}, \bibinfo {author}
  {\bibfnamefont {U.}~\bibnamefont {Schollw\"ock}}, \bibinfo {author}
  {\bibfnamefont {J.}~\bibnamefont {Eisert}}, \ and\ \bibinfo {author}
  {\bibfnamefont {I.}~\bibnamefont {Bloch}},\ }\href@noop {} {\bibfield
  {journal} {\bibinfo  {journal} {Arxiv preprint arXiv:1101.2659}} (\bibinfo
  {year} {2011})}\BibitemShut {NoStop}%
\bibitem [{\citenamefont {Lieb}\ and\ \citenamefont
  {Liniger}(1963)}]{lieb:1963}%
  \BibitemOpen
  \bibfield  {author} {\bibinfo {author} {\bibfnamefont {E.}~\bibnamefont
  {Lieb}}\ and\ \bibinfo {author} {\bibfnamefont {W.}~\bibnamefont {Liniger}},\
  }\href@noop {} {\bibfield  {journal} {\bibinfo  {journal} {Physical Review},\
  }\textbf {\bibinfo {volume} {130}},\ \bibinfo {pages} {1605} (\bibinfo {year}
  {1963})}\BibitemShut {NoStop}%
\bibitem [{\citenamefont {Tracy}\ and\ \citenamefont
  {Widom}(2008)}]{tracy:2008}%
  \BibitemOpen
  \bibfield  {author} {\bibinfo {author} {\bibfnamefont {C.}~\bibnamefont
  {Tracy}}\ and\ \bibinfo {author} {\bibfnamefont {H.}~\bibnamefont {Widom}},\
  }\href@noop {} {\bibfield  {journal} {\bibinfo  {journal} {Journal of Physics
  A: Mathematical and Theoretical},\ }\textbf {\bibinfo {volume} {41}},\
  \bibinfo {pages} {485204} (\bibinfo {year} {2008})}\BibitemShut {NoStop}%
\bibitem [{Note1()}]{Note1}%
  \BibitemOpen
  \bibinfo {note} {`Never impose on others what you would not choose for
  yourself.' -- Confucius}\BibitemShut {NoStop}%
\end{thebibliography}

%merlin.mbs 2010-03-15 4.21a (PWD, AO, DPC)

\end{document}